# Calculation of Band Structure Using Local Sampling and Green's Functions

Milad Khoshnegar, Sina Khorasani, Amirhossein Hosseinnia

*School of Electrical Engineering, Sharif University of Technology*

*P. O. Box 11365-9363, Tehran, Iran*

*Email: khorasani@sina.sharif.edu*

A new method for calculation of band structure has been proposed based on the Green's function theory and local sampling. Potential energy in the Hamiltonian of Schrödinger's equation is approximated with a series of sampled Dirac delta functions weighted by appropriate factors. These factors are found from multipole expansion of atomic potentials in the crystal lattice, with considering effects such as screening. Fourier transform was then applied to describe the wave function in reciprocal space. Sampling can be uniform or non-uniform throughout space; however rate and interval optimization is essential. Theory was implemented for Silicon, Germanium and Graphene sheet individually, while results were compared with the ab-initio non-local pseudopotential (AINLPS) method. Also for Silicon, the pseudopotential used in orbital-free density functional theory (OF-DFT) was employed as a suitable sampling source. Phase variations of the dispersion formula are analyzed, introducing adapting parameters to improve compatibility with ab-initio results. Local analysis with low order truncation in real space, reduces implementation time while giving acceptable results.

# I. INTRODUCTION

Rapid development of novel nano-structured materials demands new methods of obtaining the energy band structures, while preserving accuracy and efficiency. Extending applications of semiconductor crystals in nano-scale electronics and their dual photonic [1] and phononic [2,3] structures are enforcing to provide new and efficient methods to acquire band structures in order to characterize phenomena such as particle transport or band-to-band transitions, new dispersion effects, etc. In particular, there is a need to provide insight into novel nano-structures such as the two-dimensional (2D) Graphene [4,5] and Graphane [6], one-dimensional (1D) Carbon nanotubes and the zero-dimensional (0D) fullerene and Carbon nanotori [7-9].

At present, numerous methods for extraction of electronic band structure are in use. There exist successful methods such as density functional theory (DFT) and Quantum Monte Carlo (QMC) methods. Although DFT has the ability to simulate precise charge density [10], and much more efficient with respect to other ab-initio methods, it may still be regarded as a relatively complicated approach. Thus over the past decades analytical and numerical methods for calculating the energy dispersion have been developed; these include the free electron approximation (nearly-free electron method) [11], cellular method [12], tight-binding method [13,14], Wentzel-Kramers-Brillouin (WKB) approximation [15], scattering matrix method, QMC [16], DFT [17-19], Hartree-Fock-Slater approximation [20], and finally the so-called empirical pseudopotential method (EMPS) [21]. DFT calculations are normally equipped with the augmented plane wave [22-24] representation of orbitals. In nearly all circumstances, the adiabatic or the so-called Born-Oppenheimer approximation [25,26] is used, which allows neglecting the momentum of heavy nuclei. Also wavelet-based methods are introduced in photonic to calculate band structure of 2D photonic crystals [27,28]. This is while envelope function and $\mathbf{k}\cdot\mathbf{p}$ perturbation methods are widely used for design and analysis of quantum well [29] and dot devices [30].

In this paper, we present a new method for calculation of the electronic band structure, based on the combination of one-particle Green's function method and sampling from the electronic potential. Sampling is done in the physical space by using Dirac's delta functions as the basis of expansion in the approximation of electronic potential. The sampled effective potential can be simply the atomic potential with Coulomb screening, while the effects of the nearest neighbors are included. We have however noticed that for some materials this construction of effective potential could be inappropriate. In such cases, the ab-inito local pseudopotentials (AILPS) (or other sources as discussed later) may be used as an external input for the electronic potential to the code. Results of both models are compared with the non-local pseudopotential including spin-orbit interaction. When the atomic potentials are used subject to screening, more than one fitting parameter may be necessary for reproduction of accurate results. Furthermore, the series expansion of the one-particle wave function must be truncated; these need tuning like what is done in EMPS. However, only incorporation of the first few terms turns out to be usually sufficient.

In 3D problems, plane wave methods lead to multi-dimensional matrices with orders greater than two, which need addressing by typically more than two integer indices. For the purpose of numerical computations, it is common to map these matrices onto square matrices with the order of two, which result in very large dimensions and diminished efficiency of energy eigenvalue extraction. The problem is overcome here, by increasing sampling rate and decreasing truncation number; this leads to a more local analysis of the lattice, where eventually only one primitive cell is sampled as the building block of periodic system. This approach has also the capacity of employing the spatial potential obtained through other approaches, such as ab-initio tables which give discrete data of charge density.

We calculate the band structures of Si, Ge and graphene sheet by using the proposed method. For Si, the AILPS result was used, which is a modification of Kohn-Sham theory. We observed the errors to decrease in comparison to Ge, while truncation number was held small. For the graphene, sensitivity due to variation of delta function weights, sample locations and adapting parameters are assessed. Results are comparable with ab-initio and third-neighbor tight binding curves.

The organization of this paper is as follows: In §2 we discuss the method used to sample the potential in real space. Then in §3 we proceed to describe the expansion of Bloch waves and the Green's function approach. The next Section presents the analytical derivation of dispersion equation. Results and example test cases are presented in §5 and finally the conclusions are drawn. Furthermore, two appendices present a theorem on Dirac deltas and also discuss a method for the approximation of error in evaluation of potential coefficients.

## II. POTENTIAL SAMPLING

Approximation of the energy subbands is directly connected to the assessment of the eigenvalues for the well-known Schrödinger's equation [31]

$$(\frac{\hbar^2}{2m}\nabla^2 + \xi)\Psi_{\kappa}(\mathbf{r}) = V(\mathbf{r})\Psi_{\kappa}(\mathbf{r}), \qquad (1)$$

where $\Psi_{\kappa}(\mathbf{r})$ is the probability wave, $V(\mathbf{r})$ represents the potential energy operator and $\xi$ denotes the total energy. The electronic potential operator $V(\mathbf{r})$ has a crucial role in our analysis and needs to be appropriately dealt with, otherwise inaccurate input matrices in the computation are expected. Usually, potential well of each closed shell in crystal, especially for ionic structures, is modeled by an effective potential magnitude. This effective value, weights a delta function located at a definite point, where the nucleus or closed shell pointing vector resides [32]. This model needs simple simulation steps and yields satisfactory results, even in comparison to precise approaches such as ab-initio. This holds true in particular for substances which constitute weak orbital overlaps, e.g. Graphene [4,5], where overlap integrals of tight-binding method are small. Also the arrangement of atoms in lattice sites and their potential effect upon each other is inferior.

*2.1. Closed shell potential well*

One needs to have the electronic potential and charge density in a crystal at hand prior to sampling. The inherent difficulty of determining the correct shape of potential and charge distributions lies within the fact that their exact forms are not known unless a self-consistent solution of Poisson's and Schrödinger's equations has already been performed; but this means that the band structure should already have been available before. Hence, only an externally provided non-self-consistent, yet careful, choice of potential can lead to a more efficient computation than methods such as DFT.

To this end, several choices for a suitable model potential may be realized by either integrating the Poisson's equation based on an assumed form of charge distribution, or directly choosing a prescribed form for the effective potential. Prescription of an inaccurate charge density would result at best in a partially screened Coulomb potential. It is also possible to use the potential resulting from Wang-Parr's algorithm [33]. However, this method is iterative and too difficult to be practical for the periodic structures of our interest. The other alternative is to employ any of the available pseudopotentials [10], which replace the the strong Coulomb potential of the nucleus and surrounding electrons with an effective potential. The radial dependence of pseudopotentials is in such a way that they tend to behave like the Coulomb potential of an ionic core at large distances away from the nucleus, but deviate significantly from the $1/r$ dependence at shorter distances. AILPS and ab-initio non-local pseudopotential (AINLPS) are known to give rise to satisfactory consequences. There exist also empirical pseudopotentials, which receive data for some fitting parameters from experiments. It is noteworthy to mention that there exist models, such as the Generalized Exponential Cosine-Screened Coulomb (GECSC) potential [34], which includes an exponentially collapsing profile with taking effective interaction in many-body environment. GECSC is generally based on dividing potential into shape-invariant and perturbed terms. Then bound state energies and the corresponding wave functions are estimated for a definite perturbation order.

Charge density and potential contours are also available by ab-initio output tables, but they do not provide the suitable database in some cases. For example, theoretical charge density of Si is normally investigated through local density approximation (LDA) and generalized gradient approximation (GGA) methods. Zuo et. al. [35] performed a through Hartree-Fock theoretical study of charge density of Silicon. They compared the theoretical values with the exchange and correlation potentials constructed from the experimental data on the structure factors of Silicon, and observed good agreement (c.f. §2.2). However, this does not imply that they can be readily applicable to another method. In addition, such kinds of analyses are not available for a wide range of materials.

So we need to take on an initial yet simple sampling pattern, then we introduce some tuning factors to reproduce the appropriate potential profile. For this purpose, we expand the electrostatic potential on the spherical multipoles as

$$\Xi_0^j(\mathbf{r}) = \frac{1}{\varepsilon_{\text{sem}}} \sum_{\ell} \sum_{\mu=-\ell}^{\ell} \frac{1}{2\ell+1} \chi_{\ell\mu} \frac{Y_{\ell\mu}(\theta,\varphi)}{\left|\mathbf{r}-\mathbf{r}_j\right|^{\ell+1}}, \qquad (2)$$

where $\ell, \mu \in \mathbb{N}$, and $R_{\text{effective}} > a_0$ corresponds to an effective atomic radius larger than the Bohr radius $a_0$. $\Xi_0(\mathbf{r})$ is the spatial electrostatic potential due to $j$th atom. Also, $Y(\theta,\varphi)$ denote the spherical harmonics, and $\varepsilon_{sem}$ is the spatial average of material permittivity. It should be noted that since a background permittivity is included by incorporating $\varepsilon_{sem}$, then the screening effect is not totally ruled out; part of its contribution has actually been included in an averaged fashion. $\chi_{\ell\mu}$ coefficients would then be [36]

$$\chi_{\ell\mu} = \int \rho(\mathbf{r}')Y_{\ell\mu}(\theta',\varphi')r'^{\ell}\mathrm{d}^3 r', \tag{3}$$

in which $\chi_{\ell\mu} = (-1)^\mu \chi_{\ell\mu}^*$ and $\rho(\mathbf{r})$ is the charge density operator. The corresponding operator $\rho(\mathbf{r})$ in the case of a single closed shell and its valence electrons is

$$\rho(\mathbf{r}) = Q_{nuc}\delta(\mathbf{r}-\mathbf{r}_{nuc}) - \sum_i e\,\delta(\mathbf{r}-\mathbf{r}_{nuc}-\mathbf{r}_{ie}), \tag{4}$$

where $Q_{nuc} \in \mathbb{R}^+$ is the effective nucleus charge and $\mathbf{r}_i, i \in \mathbb{N}$, denotes the position vector of electrons measured with respect to the nucleus. It is a common practice to consider a screened Coulomb potential such as those utilized in some pseudopotential models [34,37]. Conventional screenings in plasmons or electron gas are typical examples of deviation from the primitive $1/r$ potential model. In our case, electrons around the nucleus may contribute significantly to this effect, specifically for metals. Using Thomas-Fermi formula for $\tilde{\varepsilon}(\mathbf{\kappa})$ (dielectric static function in $\mathbf{\kappa}$ space), which relates the Fourier transform of spatial charge to the generalized electrostatic potential $\tilde{\Theta}(\mathbf{\kappa})$ we have

$$\tilde{\Theta}(\mathbf{\kappa})(\kappa^2 + k_s^2) = \rho(\mathbf{\kappa}), \quad k_s = 4(3n_0/\pi)^{1/3} a_0^{-1}, \tag{5}$$

or

$$-\nabla^2\Theta(\mathbf{r}) + k_s^2\Theta(\mathbf{r}) = \rho(\mathbf{r}), \tag{6}$$

in which $k_s$ is referred to as the Thomas-Fermi screening length. Also, $n_0$ is the intrinsic carrier concentration which is dependent on the temperature. We furthermore define $\mathbf{D}(\mathbf{r}) = -\nabla\Theta_0(\mathbf{r})$. Green's function of the above differential equation, which also includes screening effect, takes the form

$$T(\mathbf{r},\mathbf{r}') = \frac{1}{|\mathbf{r}-\mathbf{r}'|}\exp(-k_s|\mathbf{r}-\mathbf{r}'|), \tag{7}$$

So, the generalized electrostatic potential is obtained with the convolution of $T(\mathbf{r},\mathbf{r}')$ and $\rho(\mathbf{r})$ as

$$\Theta(\mathbf{r}) = \rho(\mathbf{r}) \otimes T(\mathbf{r},\mathbf{r}'). \tag{8}$$

This may be expanded as:

$$\Theta(\mathbf{r}) = \varepsilon_{\text{sem}}\Xi(\mathbf{r}) = \rho(\mathbf{r}) \otimes \sum_{\ell}\sum_{\mu=-\ell}^{\ell}[\frac{4\pi}{2\ell+1}\frac{r_<^\ell}{r_>^{\ell+1}}Y_{\ell\mu}^*(\theta',\varphi')Y_{\ell\mu}(\theta,\varphi)\exp(-k_s|\mathbf{r}-\mathbf{r}'|)], \tag{9}$$

where $r_< = \min(r,r')$ and $r_> = \max(r,r')$. This expression combines both the multipole model and screening effect for an average spherical charge density.

In the case of our proposed method, however, the potential energy due to atoms is given by:

$$V(\mathbf{r}) = \frac{1}{2}\sum_j \int_{r'<r_j} \rho(\mathbf{r}-\mathbf{r}')\Xi_0^j(\mathbf{r}')\mathrm{d}^3r', \tag{10}$$

Obviously it is not usually possible to find a closed-form expression for $V(\mathbf{r})$ except in some particular cases, such as point charge assumption for $\rho(\mathbf{r})$. So one may choose to estimate (10) numerically at specific locations, where sampling takes place. We later employ this method for the 2D Graphene and 3D Ge. This method, however, fails to reproduce convincing results for some materials, where other models must be borrowed from more precise methods, like what is employed for Si in §6. The local pseudopotential (LPS) obtained by LDA assumes that the total crystal potential is given by the sum of ionic potentials estimated at a definite point (an assumption which we used above too). But the profile must also be modified to contain proper deviations from the simple $1/r$ model. In LPS, electrostatic potential is separated into estimation of two series which sweep real and reciprocal domain (Fourier space). These series must be evaluated through Ewald summation routine to care of the convergence [38].

While LPS is not convincing for structures where core-valence exchange and correlation are significant, but it can describe core-valence interactions for variational analysis. Also LPS should be enhanced to non-local pseudopotential for approaches which consider moving particles such as Monte Carlo [15]. However for single-particle simulations where many-body effects is usually neglected, LPS still works fine.

*2.2. Many-body effects*

The nuclei of the electronic lattice contribute to a static potential which shows itself in Hamiltonian operator for all $N_e$ electrons in the system. In this context, DFT is a powerful method in description of interacting electrons under influence of such potential. Also, for bulk crystals the Kohn-Sham density functional theory (KS-DFT) modified by Wang-Parr's approach [33] leads to stable computations.

In the standard DFTs, the effective single-particle potential is defined as a partial derivative of energy functional with respect to the charge density, and the exchange-correlation potentials or many-particle interactions, are added to the Coulomb repulsion. Generally, non-interacting kinetic energy contains much larger magnitudes than the exchange and correlation energies. So with utilizing KS-DFT, one can describe the system with a set of $N_e$ one-electron orbitals, while they fulfill $N_e$ coupled KS equations. In Wang-Parr's modification of this method, a self-consistent procedure is introduced to characterize the effective, through an iterative combination of LPS with KS-DFT.

Both reciprocal and real space AILPS charts due to KS-DFT approach are achieved, especially for Si [35,39,40]. In fact, one-particle KS-DFT eigenvalues are widely used in band structure calculations. These eigenvalues, which conventionally appear in KS formulation, fulfill an orthonormality condition with no apparent physical significance. This means that, if we were able to estimate the aforementioned exchange-correlation potential, there would be no justification in explaining of eigenvalues as being exact energy quantities to add or extract an electron to or from system. This yields that DFT energy gaps are normally lower than experimental amounts. So even if DFT results are used for sampling, still some fitting parameters should be imposed for precise results; this is because one would normally expect that those problems associated with DFT, would be also inherited by the present approach.

Practically, Coulomb DFT is known to have reasonable elicitation of average charge density; this is because of the dominant contribution given by Hartree-Fock in contrast to correlations (perhaps these correlations can be used as corrections, however not in the case of influential electron-electron interactions). Since discussion about features of this theory is not the main motivation here, only the relevant results will be considered in §6.2.

*2.3. Sampling approach*

We approximate the effect of the electronic potential by a three-dimensional (3D) array of weighted 3D Dirac's delta functions $\delta^{(3)}(\mathbf{r})$. A similar idea has been earlier used by Hsu and Reichl [32], who employed regularized delta functions in place of the crystal potential; they replaced each atom with one delta function and observed reasonable results for graphene and carbon nano-tubes. Instead of one sample per atom, we take many samples from the effective potential inside the unit cell to reproduce the potential with more accuracy. It is possible to use both uniform or non-uniform grids. For addressing sampling points, Cartesian and polar systems are exchangeable, however, a uniform sampling with perpendicular or radial intervals would provide different regimes of phase variations. As it will be discussed, optimization of grid size will be necessary to ensure a high degree of accuracy; this leads to definition of proper fitting parameters. Weights are simply the value of the effective potential at the points of sampling, but they can scale altogether according to a constant scaling factor, to be discussed below. The Dirac's delta functions $\delta^{(3)}(\mathbf{r})$ have a flat spectrum, but in the reciprocal space, the wavefunction is fortunately multiplied by the Fourier transform of the Green's function, which decays quickly by increasing distance from the center of the reciprocal space. Hence, in practice we only need to limit the number of summation terms, which are being taken into account.

To ensure an acceptable and fast simulation process, four fitting parameters are defined: the total number of samples (ns), the adjustment factor (AF), the interval number (N), and the scaling factor (SF). The *adjustment factor* (AF) distinguishes the boundaries of sampling domain, here referred to as the S-domain. We define the AF as the ratio of the physical width of the S-domain to the lattice constant; in practice, the AF does not exceed 0.5. The AF tunes the density of the samples (referred to as the resolution), intervals and the exceeding phase. The *interval number* (N) denotes the number of intervals along each coordinate. Hence, *ns* is a function of the AF, N, as well as the geometrical shape of primitive cell. Generally speaking, we evaluate the AF to enforce the S-domain not to overshoot cell region, then AF and N together would determine the resolution. The *scaling factor* (SF) increases or decreases all sample weights simultaneously, these weights are solely dependent on the potential profile.

The sampling domain is same as the primitive cell, which constitutes the whole crystal through the translation vectors

$$\boldsymbol{\alpha} = m\mathbf{a}_p^1 + n\mathbf{a}_p^2 + p\mathbf{a}_p^3, \tag{11}$$

in which $m, n, p \in \mathbb{N}$ are limited by simulation boundaries or crystal dimensions, and $\mathbf{a}_p^1, \mathbf{a}_p^2, \mathbf{a}_p^3$ represent primitive cell's basis vectors. In each cell, we can arbitrarily choose sampling points, but it is recommended to select samples which are located within appropriate intervals. This method helps us to grasp a reasonable image of the actual profile, based on the number of closed shells in each cell. Additionally, locations of closed shells measured from the origin are denoted by $\mathbf{S}_p^j$, where the index $j$ sweeps the number of atoms in the cell. We select the position of origin, in such a way to produce the highest possible degree of symmetry, to simplify the coding process. Normally, it is suitable to locate the origin of system at the center of the cell. Finally by introducing the destination vector $\mathbf{R}_p^{ns}$, local sampling points around the nuclei are addressed, where $ns$ distinguishes the sample index.

The necessity of regularizing the location of samples is connected to the expression for dispersion equation (c.f. §4). Expectedly, connecting vector from origin to the sampling point is the sum of two connection and destination vectors plus periodicity vector:

$$\hat{\mathbf{r}}_{ns} = \hat{\mathbf{r}}_{j\alpha} + \mathbf{R}_p^{ns} = \boldsymbol{\alpha} + \mathbf{S}_p^j + \mathbf{R}_p^{ns}, \tag{12}$$

Non-uniform sampling can be then utilized to collect data from regions with high magnitude or sharp gradients in electronic potential, which are generally distant from atoms containing less information than the nearby points. Continuous-time-domain methods for linear systems are available which anticipate abrupt changes and appropriate intervals [41]. Such methods could be exploited only for 1D problems, where an analogy to the time-coordinate exists, however for higher-dimensional structures these issues must be fully resolved.

### III. BLOCH WAVE EXPANSION AND GREEN'S FUNCTION

Application of Green's functions in calculation of energy bands, although already known, but has not been fully developed to its full capacity. Nonsingular Green's functions, extracted with no restriction on crystal potential, are common studies performed in this case [42]. Here, we introduce the idea of merging Green's function formalism with delta-based sampling, and is successfully done for the first time to our knowledge. By the Lippman-Schwinger equation the whole wave function divides into the summation of two parts: (a) free solution from Schrödinger's equation, and (b) the retarded Green's function solution obtained from convolution with the product of the wave function and the external potential. Here, we formulate the Green's function in the reciprocal domain, while in real space boundary conditions must be satisfied with respect to Dirichlet or Newmann restrictions.

It is known that the Bloch wave functions as the eigenfunctions of the periodic real space, are proportional to product of a linear phase term by a periodic function (eliminating scattering effect), as

$$\Psi_{\kappa}(\mathbf{r}) = \exp(-i\boldsymbol{\kappa}\cdot\mathbf{r})\Phi_{\kappa}(\mathbf{r}), \tag{13}$$

$$\Phi_{\kappa}(\hat{\mathbf{r}}_{ns}) = \Phi_{\kappa}(\boldsymbol{\alpha} + \mathbf{S}_{p}^{j} + \mathbf{R}_{p}^{ns}) = \Phi_{\kappa}(\mathbf{S}_{p}^{j} + \mathbf{R}_{p}^{ns}). \tag{14}$$

The aforementioned periodicity, however, does not include just potential term, but the entire Hamiltonian. Please note that the treatment of propagating Bloch waves in mass dependent media, like superlattices, is different and not presented here. Usually, envelope function approximation together with **k·p** perturbation approaches are used in such cases [29,30]. Now we express the Schrödinger's equation in Rydberg units

$$\nabla^2 \Psi_{\kappa}(\mathbf{r}) + \xi\, \Psi_{\kappa}(\mathbf{r}) = \sum_{\boldsymbol{\alpha}} V_{\mathrm{p}}(\mathbf{r})\, \Psi_{\kappa}(\mathbf{r}), \tag{15}$$

where $V_{\mathrm{p}}(\mathbf{r})$ is related to the potential distribution in the sub-domains or primitive cells (index $p$ pertains to primitive). For the Bloch wave function, (15) appears with an imaginary shift in gradient

$$[(\nabla - i\boldsymbol{\kappa})^2 + \xi]\Phi_{\kappa}(\mathbf{r}) = \sum_{\boldsymbol{\alpha}} V_{\mathrm{p}}(\mathbf{r})\, \Phi_{\kappa}(\mathbf{r}). \tag{16}$$

We can rewrite (16) using delta function expansion of $V_{\mathrm{p}}$

$$V_{\mathrm{p}}(\mathbf{r}) = \sum_{j=1}^{\mathcal{L}} \sum_{ns=1}^{\mathrm{N}} C_{ns} \delta^{(3)}(\mathbf{r} - \hat{\mathbf{r}}_{ns}), \tag{17}$$

where $\mathcal{L} \in \mathbb{N}$ and $\mathrm{N} \in \mathbb{N}$ denote total number of atoms and sampling points in a sub-domain, respectively, and $C_{ns} \in \mathbb{R}$ is the normalized coefficient corresponding to potential value in the appropriate point. It is evident that we can reconstruct (16) based on the Green's function theory in the form of

$$\Phi_{\kappa}(\mathbf{r}) = -G_{\kappa}(\mathbf{r},\mathbf{r_0}) \otimes [V(\mathbf{r})\Phi_{\kappa}(\mathbf{r})], \tag{18}$$

where $G_{\kappa}(\mathbf{r},\mathbf{r_0})$ satisfies

$$[(\nabla - i\kappa)^2 + \xi]G_{\kappa}(\mathbf{r},\mathbf{r_0}) = -\delta^{(3)}(\mathbf{r}-\mathbf{r_0}). \tag{19}$$

This shows that the impulse response of a system with the Hamiltonian $\hat{H} = \{(\nabla - i\kappa)^2 + \xi\}$ is $G_{\kappa}(\mathbf{r},\mathbf{r_0})$.

Thus, the response of a similar system with the same Hamiltonian to $V(\mathbf{r})\Phi_{\kappa}(\mathbf{r})$ is given by

$$\Phi_{\kappa}(\mathbf{r}) = G_{\kappa}(\mathbf{r},\mathbf{r_0}) \otimes \sum_{\alpha'}\sum_{j=1}^{L}\sum_{ns=1}^{N} C_{ns}\delta^{(3)}(\mathbf{r}-\hat{\mathbf{r}}_{ns})\Phi_{\kappa}(\mathbf{r}), \tag{20}$$

$$\alpha' = m\mathbf{a}_p^1 + n\mathbf{a}_p^2 + p\mathbf{a}_p^3 \quad m_1, n_1, p_1 \in \mathbb{N}, \tag{21}$$

Now we substitute Fourier series expansion as the basis of $\Phi_{\kappa}(\mathbf{r})$ in (20). This results in

$$\Phi_{\kappa}(\mathbf{r}) = \sum_{\alpha}\tilde{\Phi}_{\alpha}\exp(i\mathbf{\Gamma}_{\alpha}.\mathbf{r}) = -G_{\kappa}(\mathbf{r},\mathbf{r_0}) \otimes \sum_{\alpha'}\sum_{j=1}^{L}\sum_{ns=1}^{N}[C_{ns}\delta^{(3)}(\mathbf{r}-\hat{\mathbf{r}}_{j\alpha} - \mathbf{R}_p^{ns})\Phi_{\kappa}(\mathbf{S}_p^j + \mathbf{R}_p^{ns})], \tag{22}$$

$$\mathbf{\Gamma}_{\alpha} = 2\pi(m\overline{\mathbf{b}}_r^1 + n\overline{\mathbf{b}}_r^2 + p\overline{\mathbf{b}}_r^3), \tag{23}$$

where $\mathbf{\Gamma}_{\alpha}$ denotes the transfer vector and $\overline{\mathbf{b}}_r^1, \overline{\mathbf{b}}_r^2, \overline{\mathbf{b}}_r^3$ represent the conventional basis vectors in the reciprocal lattice. Due to the periodicity of $\Phi_{\kappa}(\mathbf{r})$, its Fourier transform could be achieved simply with respect to the expansion coefficients.

## IV. DISPERSION EQUATION

Fourier transform of the periodic envelope function of the Bloch wave can be found by summing Fourier series coefficients multiplied by the corresponding delta functions in reciprocal space:

$$\mathcal{F}\{\Phi_{\kappa}(\mathbf{r})\}(\mathbf{\eta}) = (2\pi)^3\sum_{\alpha}[\tilde{\Phi}_{\alpha}\prod_{\nu=1}^{N=3}\delta(\eta_{\nu} - \mathbf{\Gamma}_{\alpha}.\mathbf{a}_p^{\nu})]. \tag{24}$$

Exploiting this transformation in (22) gives

$$(2\pi)^3\sum_{\alpha}[\tilde{\Phi}_{\alpha}\prod_{k=1}^{N=3}\delta(\eta_k - \mathbf{\Gamma}_{\alpha}.\mathbf{a}_p^k)] = -\tilde{G}_{\kappa}(\mathbf{\eta},\mathbf{r_0}) \times \sum_{\alpha'}\sum_{j=1}^{L}\sum_{ns=1}^{N} C_{ns}\Phi_{\kappa}(\mathbf{S}_p^j + \mathbf{R}_p^{ns}) \times \exp[-i\mathbf{\eta}.(\alpha' + \mathbf{S}_p^j + \mathbf{R}_p^{ns})], \tag{25}$$

in which $\tilde{G}_{\kappa}(\mathbf{\eta},\mathbf{r_0}) = \mathcal{F}\{G_{\kappa}(\mathbf{r},\mathbf{r_0})\}$. Referring to (19), $\mathcal{F}\{G_{\kappa}(\mathbf{r},\mathbf{r_0})\}$ is calculated as

$$\tilde{G}_{\boldsymbol{\kappa}}(\boldsymbol{\eta},\mathbf{r_0}) = \frac{-\exp(-i\,\boldsymbol{\eta}.\mathbf{r_0})}{\xi - |\boldsymbol{\eta}-\boldsymbol{\kappa}|^2}, \tag{26}$$

$$\lim_{\tau\to\infty}\int_{-\tau}^{\tau}|G(\mathbf{r},\mathbf{r_0})|^2\,d^3r = \lim_{\omega\to\infty}(2\pi)^{-3}\int_{-\omega}^{\omega}|G(\boldsymbol{\eta},\mathbf{r_0})|^2\,d^3\eta < \infty \tag{27}$$

With a realistic assumption for energy eigenvalues of a material bulk, it is convincing to have $\xi = K^2 = -k^2 < 0$, $K = ik$ (in Rydberg units). This is due to the potential wells with negative values, which enforce $C_{ns}$ coefficients to be negative, too. Note that in (25) the exponential term on the right hand side can be divided into separate expressions ($\exp[-i\,\boldsymbol{\eta}.(\mathbf{S}_p^j + \mathbf{R}_p^{ns})]$, $\exp[-i\,\boldsymbol{\eta}.\boldsymbol{\alpha}']$); and the first has no dependence on $\boldsymbol{\alpha}'$. Finally, applying the Expansion Theorem in Appendix I results in:

$$(2\pi)^3 \sum_{\boldsymbol{\alpha}}[\tilde{\Phi}_{\boldsymbol{\alpha}}\prod_{\nu=1}^{N=3}\delta(\eta_\nu - \boldsymbol{\Gamma}_{\boldsymbol{\alpha}}.\mathbf{a}_p^\nu)] = -\tilde{G}_{\boldsymbol{\kappa}}(\boldsymbol{\eta},\mathbf{r_0}).(2\pi)^3 \sum_{\boldsymbol{\alpha}'}\sum_{j=1}^{L}\sum_{ns=1}^{N}[C_{ns}\Phi_{\boldsymbol{\kappa}}(\mathbf{S}_p^j+\mathbf{R}_p^{ns})\exp[-i\,\boldsymbol{\eta}.(\mathbf{S}_p^j+\mathbf{R}_p^{ns})]\prod_{\nu=1}^{N=3}\delta(\eta_\nu + \boldsymbol{\Gamma}_{\boldsymbol{\alpha}'}.\mathbf{a}_p^\nu)]. \tag{28}$$

If $m, n, p$ and $m_1, n_1, p_1$ sweep a symmetric range such as $[-\mathcal{C}, \mathcal{C}]$, we would have delta functions on both sides while their respective weighting coefficients must equate. Also noticing the characteristics of the delta function, $\boldsymbol{\eta}$ should be replaced with $\boldsymbol{\Gamma}_{\boldsymbol{\alpha}}$ therein. This is because $\eta_1 = \boldsymbol{\Gamma}_{\boldsymbol{\alpha}}.\mathbf{a}_p^1 = 2\pi m$, $\eta_2 = \boldsymbol{\Gamma}_{\boldsymbol{\alpha}}.\mathbf{a}_p^2 = 2\pi n$, and $\eta_3 = \boldsymbol{\Gamma}_{\boldsymbol{\alpha}}.\mathbf{a}_p^3 = 2\pi p$ must hold. Hence

$$\tilde{\Phi}_{\boldsymbol{\alpha}} = -\sum_{j=1}^{L}\sum_{ns=1}^{N}C_{ns}\Phi_{\boldsymbol{\kappa}}(\mathbf{S}_p^j+\mathbf{R}_p^{ns})\tilde{G}_{\boldsymbol{\kappa}}(\boldsymbol{\Gamma}_{\boldsymbol{\alpha}},\mathbf{r_0})\exp[-i\boldsymbol{\Gamma}_{\boldsymbol{\alpha}}.(\mathbf{S}_p^j+\mathbf{R}_p^{ns})]. \tag{29}$$

The Fourier transform of the Green's function therefore is

$$\tilde{G}_{\boldsymbol{\kappa}}(\boldsymbol{\Gamma}_{\boldsymbol{\alpha}},\mathbf{r_0}) = \frac{\exp(-i\,\boldsymbol{\Gamma}_{\boldsymbol{\alpha}}.\mathbf{r_0})}{k^2 + |\boldsymbol{\Gamma}_{\boldsymbol{\alpha}} - \boldsymbol{\kappa}|^2}, \tag{30}$$

Equation (29) would be more meaningful if we rewrite $\Phi_{\boldsymbol{\kappa}}$ at sampling points, again based on the Fourier series expansion as

$$\Phi_{\boldsymbol{\kappa}}(\mathbf{S}_p^j + \mathbf{R}_p^{ns}) = \sum_{\boldsymbol{\beta}}\tilde{\Phi}_{\boldsymbol{\beta}}\exp[i\boldsymbol{\Gamma}_{\boldsymbol{\beta}}.(\mathbf{S}_p^j+\mathbf{R}_p^{ns})]. \tag{31}$$

Then

$$\tilde{\Phi}_\alpha = -\frac{1}{k^2 + |\Gamma_\alpha - \kappa|^2} \sum_\beta \sum_{j=1}^{L} \sum_{ns=1}^{N} [C_{ns} \tilde{\Phi}_\beta \exp(-i\Gamma_\alpha . \mathbf{r_0}) \exp[i\Gamma_\beta . (\mathbf{S}_p^j + \mathbf{R}_p^{ns})] \exp[-i\Gamma_\alpha . (\mathbf{S}_p^j + \mathbf{R}_p^{ns})]], \qquad (32)$$

where $\beta = m_1 \mathbf{a}_p^1 + n_1 \mathbf{a}_p^2 + p_1 \mathbf{a}_p^3$. The result would lead to a system of equations which constitutes the expansion coefficients that are assembled from (32). Within a bulk material containing no defects, the periodic part of propagating wave attains sinusoidal configuration for a specific wavevector, and this can be the first verification to examine the solution to this system of equations. Of course, this rule holds true only for single-element materials, and must be revised for alloys and compounds. Another perception is the rate of decay for $\tilde{\Phi}_\alpha$ with respect to growth of $m$, $n$ and $p$. Truncation is reliable only when out-of-range coefficients are small enough.

In the above summation, other than sample weights and Fourier series coefficients, there are linear phase terms dependent on $\mathbf{S}_p^j + \mathbf{R}_p^{ns}$, $\Gamma_\beta - \Gamma_\alpha$ and $\Gamma_\alpha . \mathbf{r_0}$. This normally results in a slight dependence of results on the absolute choice of origin, since the summations have to be truncated and imaginary parts may not completely cancel out. Hence, the sampling vector in a primitive cell plays a critical role in determination of final eigenvalues. Especially as mentioned before, the choice of the cell's origin in such a way to result in the maximally symmetric $\mathbf{S}_p^j$ vectors, is particularly beneficial if our truncation limits of $m, n, p$ and $m_1, n_1, p_1$ are symmetric or belong to $[-\mathcal{C}, \mathcal{C}]$. Normally we are able to set $\mathbf{r_0}$ at the cell's midpoint, causing zero phases as a desirable condition. On the other hand if the magnitude of destination vector $\mathbf{R}_p^{ns}$ is kept small enough, the additional phase made by $(\Gamma_\beta - \Gamma_\alpha) . \mathbf{R}_p^{ns}$ introduces a slight variation in the real part of coefficients. This additional phase, for a wide range of $\mathbf{R}_p^{ns}$ would even be comparable with the phase produced by $(\Gamma_\beta - \Gamma_\alpha) . \mathbf{S}_p^j$. This affects the real parts of the coefficients in turn. This is the case for the farther sampling points within the primitive cell in contrast to the atom's position. The adjustment factor introduced in §2.3, tunes this phase variation maxima as well as sample intervals. This adjustment factor may also be assumed as a fitting parameter to empirical results.

Now, the first summation in (32) can be decomposed into three scalar summations. Thus,

$$\tilde{\Phi}_{mnp} = -\frac{1}{k^2 + \left|\Gamma_{mnp} - \kappa\right|^2} \sum_{m_1} \sum_{n_1} \sum_{p_1} \sum_{j=1}^{\mathcal{L}} \sum_{ns=1}^{N} [C_{ns} \tilde{\Phi}_{m_1 n_1 p_1} \exp(-i\Gamma_{mnp}.\mathbf{r_0}) \exp[i(\Gamma_{m_1 n_1 p_1} - \Gamma_{mnp}).(\mathbf{S}_p^j + \mathbf{R}_p^{ns})]]. \quad (33)$$

To extract the energy dispersion equation after finding all $\tilde{\Phi}_{-\mathcal{T}^{(3)},\ldots,\mathcal{T}^{(3)}}$, we consider (13) and recast for $\Phi_\kappa(\mathbf{r})$ to obtain

$$\Psi_\kappa(\mathbf{r}) = \sum_\alpha \tilde{\Phi}_\alpha \exp[i(\Gamma_\alpha - \kappa).\mathbf{r}], \quad (34)$$

in which $\tilde{\Phi}_\alpha$ or $\tilde{\Phi}_\beta$ for all $m, n, p, m_1, n_1, p_1 \in [-\mathcal{T}, \mathcal{T}]$ are definite, and estimation of $\Psi_\kappa(\mathbf{r})$ is just restricted to a substitution. At this stage, only energy eigenvalues are found. In the next step we focused on estimating $\tilde{\Phi}_{mnp}$ coefficients to find the corresponding Bloch eigenfunctions. For this purpose, it is sufficient to rewrite (32) in the form

$$\xi.\tilde{\Phi}_\alpha = \left|\Gamma_\alpha - \kappa\right|^2 \tilde{\Phi}_\alpha - \sum_\beta a_{\alpha,\beta} \tilde{\Phi}_\beta, \quad (35)$$

where

$$a_{\alpha,\beta} = \sum_{j=1}^{\mathcal{L}} \sum_{ns=1}^{N} C_{ns} \exp(-i\Gamma_\alpha.\mathbf{r_0}) \exp[i(\Gamma_\beta - \Gamma_\alpha).(\mathbf{S}_p^j + \mathbf{R}_p^{ns})]. \quad (36)$$

In fact, (35) may be regarded as the final formula for eigenstates.

## V. RESULTS AND DISCUSSIONS

Here we present the computation results for three widely used crystals: silicon, germanium, and graphene. In all three cases, we notice good agreement with other methods.

Along with the four fitting parameters defined in §2.3, we also define the *truncation number* ($\mathcal{T}$) which determines the dimension of the final matrix. It is noteworthy that the overall computational time and the accuracy of results depend on a tradeoff among AF, *ns* and $\mathcal{T}$. We also define the *number of close neighbors* (*ncn*), which directly influences the potential profile, and thereby the sample weights. In practice, the major bottleneck which limits the computational efficiency is the rapid growth of the dimension of matrix representation of (33).

*6.1. Germanium in Diamond Structure*

For Ge with a diamond-like structure (simply achieved by combination of two displaced FCC lattices with a separation of $[a/4, a/4, a/4]$), the primitive cell can be defined by considering the FCC's cell transmitted by a $[a/8, a/8, a/8]$ transfer vector. For diamond, connection vectors are $\mathbf{S}_p^1 = [3/8, 3/8, 3/8]a$ and $\mathbf{S}_p^2 = -\mathbf{S}_p^1$. When origin is located at the midpoint of the cell, maximum phase symmetry in (33) is attained.

Here, we utilized the multipole expansion model of screened Coulomb potential for Ge (in contrast to the AILPS source for the case of Si in §6.2). Nevertheless, the results for Ge demonstrate slight deviations in most regions, even with a small truncation number. Applying a uniform sampling with AF=0.1, N=4, *ns*=125 (total 250 samples in primitive cell) and *ncn*=24, the discrete potential distribution is obtained with similar intervals such as illustrated in Fig. 1.

As mentioned, increasing *ns* with a fixed AF enhances the sensitivity of sampling to potential degree of diversity. This can be interchangeably compared to truncation number. Obviously low order truncation for a reasonable reconstruction of the local potential in the lattice calls for denser sampling. In addition, the AF must be so tuned that the sample weights also contain weak contributions from distant points away from the core. Another condition may be recognized while the magnitudes of samples are examined with realistic ab-initio results. Although deviations in $C_{ns}$ are not followed by the same scale errors in the band structure, but it may have drawbacks on the effective mass tensor. So SF variations are not allowed in a wide range.

To enhance the resolution, N was enlarged to 10 yielding 2662 ($2 \times 11^3$) samples in cell. In Fig. 2 the potential well in S-domain is shown for a Ge atom. As it is demonstrated, the profile represents the deviations from the ordinary Coulomb potential due to exponential term in (9). In Fig. 2b the AF is changed from 0.25 to 0.1. Not only the resolution is enhanced for a constant value of N=10, but also X and Y (and Z) maxima decrease. Hence, the new interval should be 0.4 times the previous one; this, guarantees the points at the depth of the well to be also sampled.

Decay regimes for series coefficients of Bloch waves are shown in Fig. 3 for $\mathcal{C}=2$. The vertical axis denotes the norm of the eigenvector, which is plotted with respect to the first five eigenvalues. The horizontal axis is the number corresponding to the variations in the indices *m*, *n* and *p* (for $\mathcal{C}=2$, 125 combinations of *m*, *n* and *p* is available), being here denoted by M. A similar regime could be obtained for $\mathcal{T}=3$ with the total number of 343. Distributions for $\tilde{\Phi}_\alpha^{1,\ldots,5}$ are symmetric relative to 62th point on horizontal axis. But what is important here, is the diminishing of maxima for M<20 and M>100. This means that we can be quite confident that for $m, n, p > O \in \mathbb{Z}$, the coefficients of (32) are adequately weak.

Ge energy bands are here extracted, accompanied by no correction, except regularizing the AF. We utilized the same profile as shown in Fig. 2. Since except the screening effect no other phenomenon was supplemented to the multipole expansion, this degree of accuracy is quite acceptable, and in fact surprising. The band structure of Ge for the first six eigenvalues is shown in Fig. 4. We note that for a low truncation number ($\mathcal{T}=3$) and matrix representation with the dimension of 343×343, noticeable deviations just occur at $\Gamma_{6c}$ and $L_{6c}$ (see Table 1). For this specific truncation number, the upper band turns aside further than five lower bands as it is obvious on the L-Λ-Γ path or X point; increasing $\mathcal{T}$ compensates these deviations. Furthermore, *ns* has a saturation limit. In other words, above the saturation limit, decreasing intervals does not correct curvature any further, and it just slows down the overall simulation. With the choice of AF=0.1 and *ns*=125, the gap is found to be about 0.74eV, which is much closer to the actual value of 0.67eV than the value predicted by N-L pseudopotential of 0.9eV. Also for an *ns* larger than 1331, the AF must be decreased to attain the same result. This is connected to the total energy gathered in the S-domain, which is tunable with AF, *ns* and SF.

Numerical data at the high symmetry points are listed in Table.1, which compares to the AINLPS including spin-orbital interaction [43]. In a simple way, we may estimate the relative error by subtracting data from two methods in Table.1, and then dividing by the average quantity. The mean value of defined error for all symmetry points is estimated to be 10.32%, to which the upper band has the largest contribution. The error collapses with the growth of truncation number; but this growth from another other point of view weakens the locality of analysis. Anyway, we can assert that applying local sampling method requires much smaller matrices than the common plane wave method, and best results are attained by $\mathcal{T}=3$.

*6.2. Silicon in Diamond Structure*

Silicon is regarded as the main material with lots of applications in the semiconductor technology, and has been extensively analyzed through the concept of atomic charge density by DFT theories. For Si, modeling of each closed shell with one unique delta function requires various adjustments to improve the fitness with pragmatic patterns. So obtaining a good sampled profile with numerous weighted delta functions needs tuning, too; the difference originates from that with a given profile, changing one delta weight leads to the scaling of all other values using SF (in the case of Ge, we used the SF as is shown in Fig. 2).

The spatial charge used here is obtained from the AILPS method. AILPS expresses the charge density with the function $t_{\rho KS}(r_c) + d_0 r^{F_1} \exp[-r^{F_2}(d_1 + d_2 r^2)]$ for a radius $r$ less than $r_c$, while it employs the standard KS model beyond $r_c$. The appropriate measures to set formula constants are given by Chai and Weeks [40].

We extended borders of S-domain to $9r_c$, nearly 3 times greater than the radius where $V_{ps}(r)$ and Coulomb potential are compatible. A total of 6776 samples were taken surrounding each atom, while truncation number was kept fixed at 3. The average error with the same definition as for Ge in §6.1, was noticed to be only 6.83% for Si, as compared to the high symmetry points of the energy dependent pseudopotential method [43].

*6.3. Graphene*

Since its discovery in 2004 [44], the 2D Graphene's honeycomb lattice as a substructure of the 1D carbon nano-tube (CNT) has received particular attention [45-48]. Ignoring the slight out of plane bending of Carbon-Carbon connection, CNT is the rolled-up Graphene plane with a definite chirality vector; also calculating its band structure is simple with zone folding method, if Graphene energy subbands are accessible through common tight binding approaches [4-6]. In addition, Graphene itself with the high degree of conductivity and anomalous Hall effect (at high magnetic fields), has the aptitude for being used in ultrafast transistors.

Sampling in the lozenge-shape primitive cell of Bilayer Graphene may also contain effects from the other layer (symmetric or asymmetric), especially when considering the C-C bond length (0.142nm from ab-initio analysis) which is comparable to the layer-layer distance. Anyway, in this paper we consider the monolayer honeycomb structure with a lattice constant of 0.242nm; this can be functional to examine the effect of background potential on $\pi$ electrons ($P_z$ orbital). To evaluate the desirable potential, again we use the screened multipole model. In addition to the first three nearest neighbors, the effects of the six second nearest neighbors effect were also brought into account or *ncn*=9 (see Fig. 7a). S-domain can be the entire lozenge-shape cell or a partial region based on the AF. We may also prefer a radial sampling, taking no care of the cross-bordering across cells. The center of the S-domain was placed amid $A_1$ and $A_2$; $\mathbf{S}_p^1$ and $\mathbf{S}_p^2$ are then given by $(a/2\sqrt{3},0)$ and $(-a/2\sqrt{3},0)$, respectively. Considering that $\mathbf{\Gamma}_\alpha$ is equal to $[(m+n).2\pi/(\sqrt{3}\,a),(m-n).2\pi/a]$, and the similarly in $\mathbf{\Gamma}_\beta$ subscripts changes as $[(m_1+n_1).2\pi/(\sqrt{3}\,a),(m_1-n_1).2\pi/a]$, the produced phase by $(\mathbf{\Gamma}_\beta - \mathbf{\Gamma}_\alpha).\mathbf{S}_p^j$ would be obtained explicitly as $\pi/3(m+n-m_1-n_1)$.

Here we employed a radial sampling. Paths of sampling supposed to be circles, while parameters of the polar system (radius and angle) had similar intervals individually. Now, let us discuss about role of thAF in controlling the excess phase, because we found out that $\pi$ bands of Graphene are so sensitive to this excess phase. At first, the AF tunes $\left|\mathbf{R}_p^{ns}\right|$ maxima. When $\mathbf{\Gamma}_\beta - \mathbf{\Gamma}_\alpha$ and $\mathbf{R}_p^{ns}$ have approximately close directions, $\left|\mathbf{R}_p^{ns}\right|$ contribute to an important effect on the phase variations of $a_{\alpha,\beta}$. This may suggest that with a smaller AF and constant *ns*, the maximum range of the excess phase would decrease. On the other hand, the excess phase depends on the exact location of samples, and we know with a constant N (*ns*), the AF specifies this location. Because of this, we would see in the following that a slight change in the AF affects the eigenvalues dominantly, which makes the calculation of Graphene's band structure by this method very difficult. It is notable that for a symmetric sampling, the imaginary parts of $a_{\alpha,\beta}$ cancel out exactly; so the AF just affects the real part of $a$ for a definite **α, β**. The excess phase for two different the AF is shown in Fig. 8. For comparison, we tuned *ns* in each to result in nearly equal sample density. It is shown that for a larger S-domain, the maximum excess phase is greater. In addition, although *ns* is almost the same, but phase diagram for AF=0.5 exhibits stronger oscillations. In §6.2 for Si, we preferred not to use the AF as an adjustment factor, because our purpose was to extend the S-domain until Coulomb model and $V_{ps}(r)$ become compatible; however in the case of Ge, the choice of AF=0.2 was noticed to be the optimum value.

To examine our model on Graphene, we extracted the two energy bands ( $\pi$ bands) near the Fermi level. The triangular half of the lozenge was considered as an independent S-domain. With AF=0.38 and $\mathcal{T}=10$, we changed delta weights smoothly to gain the potential profiles shown in Fig. 9b. If we notice that the first three neighbors are located at $(a/\sqrt{3},0)$, $(a/\sqrt{3},2\pi/3)$ and $(a/\sqrt{3},4\pi/3)$ for $A_1$, a potential fall occurs distant from core for angle=0. This embodiment mirrors for $A_2$. For the regularized AF, SF was altered from 0.8 to 1.25. Here, SF=1 was noticed to give close values to the ab-initio results ($V_{app}$ denotes the best responding profile). Enlarging SF yields a decrease of mid-gaps at $\Gamma$ and M. In Fig. 9c, the corresponding eigenvectors are shown. For $\mathcal{T}=10$, 441 combinations of $m$, $n$ and $p$ are available and here just 250 vectors with largest maxima are shown.

As expected, increasing delta weights results in small changes at M, but deviates $\Gamma$ considerably. Comparison to the ab-initio method is shown in Table 2 at high symmetry points. During simulation it was found that the sensitivity of $\pi$ bands to truncation numbers over 10 and for this special AF was low. Consequently, improvement of results only with imposing larger truncation would not be achieved, although not impossible with another SF and AF.

On the other hand, the sensitivity to AF for a definite SF and truncation is noticeable. This is clearly shown in Fig. 10a, where for similar potential, truncation and sample intervals, the conduction and valence bands are plotted with changing AF from 0.33 to 0.38. The excess phase was calculated in an independent process where all samples were swept for $m, n, m_1, n_1 \in [-1,1]$. We elicit although variation of AF is so slight, but conduction band reveals a large alteration relatively. This is so, because sample weights contribute to S-domain (consider sample intervals are alike and changes of excess phase are small comparatively); practically, the effective potential or the mean value of weights decreases for a lower AF. This sensitivity usually becomes apparent on $\Gamma_c$, where the magnitude of wavevector approaches to zero.

# VI. CONCLUSION

We have devised a new method based on the combination of Green's function formalism and local sampling by Dirac delta functions in a periodic lattice. We have shown that the method may be expected to produce satisfactory results. The discussed approach shows superior performance, accuracy, and efficiency over most of the other existing numerical and analytical approaches. The method has been tested against the well-studied 3D and 2D crystals, namely Silicon, Germanium, and Graphene. In all cases, reasonable agreement to other approaches were noticed. For the case of Graphene, the size of series expansion did not exceed ten, while results were close to ab-initio and somewhat superior to the third-neighbor tight-binding. For the 3D lattices (Ge and Si) with a high order density of samples we were able to keep the truncation number below five, while maintaining good accuracy. Although precise computation of the energy bands still require greater truncations to compete with ab-initio technique. We note that the algorithm was interiorly responsive to the selection of the S-domain, intervals and in particular, the sampling. Further extension of this work is needed to include spin-orbit interaction, which will be hopefully the subject of a separate study.

# ACKNOWLEDGEMENT

The authors wish to thank the reviewers of this work for providing constructive comments.

# APPENDIX I. EXPANSION THEOREM

Here, we present a theorem on the series of delta function, which is needed for expansion of (22).

*Theorem.* Let $\mathbf{\Gamma}_{\alpha} = (\overline{\mathbf{b}}_r^1, \overline{\mathbf{b}}_r^2, \overline{\mathbf{b}}_r^3)$ be the reciprocal space transferring vector and $\boldsymbol{\alpha} = (\mathbf{a}_p^1, \mathbf{a}_p^2, \mathbf{a}_p^3)$ be the lattice periodicity vector. Then the completeness equation reads

$$\sum_{\alpha'} \exp(i\,\boldsymbol{\eta}.\boldsymbol{\alpha}') = (2\pi)^3 \sum_{\alpha'}[\delta(\eta_1 + \mathbf{\Gamma}_{\alpha'}.\mathbf{a}_p^1)\delta(\eta_2 + \mathbf{\Gamma}_{\alpha'}.\mathbf{a}_p^2)\delta(\eta_3 + \mathbf{\Gamma}_{\alpha'}.\mathbf{a}_p^3)] = (2\pi)^3 \sum_{\alpha'} \prod_{\nu=1}^{N=3} \delta(\eta_\nu + \mathbf{\Gamma}_{\alpha'}.\mathbf{a}_p^\nu), \quad (A.1)$$

where $\boldsymbol{\eta}$ is defined as the 3D Fourier transform vector:

$$\boldsymbol{\eta} = \eta_1 \overline{\mathbf{b}}_r^1 + \eta_2 \overline{\mathbf{b}}_r^2 + \eta_3 \overline{\mathbf{b}}_r^3 \quad (A.2)$$

*Proof.* Considering $\bar{\mathbf{b}}_r^i \cdot \mathbf{a}_p^j = 0$ for $i \neq j \in 1,2,3$ and $\eta_i = \mathbf{\eta} \cdot \mathbf{a}_p^i$ we can write down

$$\sum_{\mathbf{a}'} \exp(i\, \mathbf{\eta}\cdot\mathbf{a}') = \sum_{m_1} \exp(im_1\eta_1) \sum_{n_1} \exp(in_1\eta_2) \sum_{p_1} \exp(ip_1\eta_3) = \prod_{\substack{\nu=1 \\ l=m_1,n_1,p_1}}^{N=3} \sum_{l} \exp(i\, l\eta_\nu). \quad (A.3)$$

Through direct application of Fourier series theorem, it would be straightforward to show that

$$\sum_{\mathbf{a}'} \exp(i\, \mathbf{\eta}\cdot\mathbf{a}') = \prod_{\substack{\nu=1 \\ l=m_1,n_1,p_1}}^{N=3} [2\pi \sum_{l} \delta(\eta_\nu + 2\pi l)] = (2\pi)^3 \sum_{\mathbf{a}'} \prod_{\nu=1}^{N=3} \delta(\eta_\nu + \mathbf{\Gamma}_{\mathbf{a}'}\cdot\mathbf{a}_p^\nu). \quad (A.4)$$

This completes the proof. □

It should be mentioned that for crystals having limited physical dimensions, the above theorem is approximately satisfied if typically ten or more lattice constants across each principal direction is taken into account.

**APPENDIX II. ERROR APPROXIMATION**

In the discussion which follows, we present a systematic method for estimation of error $c_e^{ns}$ arising in the evaluation of the coefficients $C^{ns}$, where $C'_{ns} = C_{ns} + c_e^{ns}$. This is done by introducing another Green's function for the Schrödinger's equation. Suppose that the operator $\mathcal{D}$, eigenstates $\Psi_\kappa^\gamma(\mathbf{r})$ and eigenvalues $\xi_\gamma$ obey

$$\mathcal{D}\Psi_\kappa^\gamma(\mathbf{r}) = \xi_\gamma \Psi_\kappa^\gamma(\mathbf{r}), \qquad \mathcal{D} = \mathcal{T} + V, \gamma \in \mathbb{N}, \quad (II.1)$$

where $\mathcal{T}$ is again a universal operator of non-interacting kinetic energy and $V$ offers system dependent potential. For the Green's function satisfying $\mathcal{D}\Omega(\mathbf{r},\mathbf{r}_0) = \delta^{(3)}(\mathbf{r} - \mathbf{r}_0)$, we have

$$\Omega(\mathbf{r},\mathbf{r}_0) = \sum_{\gamma \in \mathbb{N}} \frac{\Psi_\kappa^\gamma(\mathbf{r}) \Psi_\kappa^\gamma{}^*(\mathbf{r}_0)}{\xi_\gamma}. \quad (II.2)$$

Here, we rearrange the operator $\mathcal{D}$ to obtain in Rydberg units to get:

$$[-\nabla^2 + \sum_{\boldsymbol{\beta}}^{\mathcal{L}} \sum_{j=1}^{N} \sum_{ns=1}^{N} C'_{ns} \delta^{(3)}(\mathbf{r} - \hat{\mathbf{r}}_{ns})] \Psi_\kappa^\gamma(\mathbf{r}) = \xi_\gamma \Psi_\kappa^\gamma(\mathbf{r}), \quad (II.3)$$

$$[-\nabla^2 + \sum_{\beta}\sum_{j=1}^{\mathcal{L}}\sum_{ns=1}^{N} C'_{ns}\delta^{(3)}(\mathbf{r} - \hat{\mathbf{r}}_{ns})]\,\Omega(\mathbf{r},\mathbf{r}_0) = \delta^{(3)}(\mathbf{r}-\mathbf{r}_0),\qquad(\text{II.4})$$

Hence, the Fourier transform of the new Green's function can be found as

$$\tilde{\Omega}(\boldsymbol{\eta},\mathbf{r}_0) = \frac{\exp(-i\boldsymbol{\eta}\cdot\mathbf{r}_0)}{|\boldsymbol{\eta}|^2 + \sum_{\beta}\sum_{j=1}^{\mathcal{L}}\sum_{ns=1}^{N} C'_{ns}\exp(-i\boldsymbol{\eta}\cdot\hat{\mathbf{r}}_{ns})} = \sum_{\gamma\in\mathbb{N}} \frac{\mathcal{F}\{\Psi_{\kappa}^{\gamma}(\mathbf{r})\}\Psi_{\kappa}^{\gamma\,*}(\mathbf{r}_0)}{\xi_{\gamma}}.\qquad(\text{II.5})$$

Here, we notice that $\boldsymbol{\eta}$ is related to the Bloch wavevector $\boldsymbol{\kappa}$ implicitly. In fact, $\tilde{\Omega}(\boldsymbol{\eta},\mathbf{r}_0)$ is not a continuous function of $\boldsymbol{\eta}$ in the reciprocal space, because of the Fourier transform of $\Psi_{\kappa}^{\gamma}(\mathbf{r})$. Now, various terms in the right hand side of (II.5) can be extracted explicitly to obtain

$$\tilde{\Psi}_{\kappa}^{\gamma}(\boldsymbol{\eta}) = (2\pi)^3 \sum_{\alpha} \tilde{\Phi}_{\alpha}^{\gamma} \prod_{\nu=1}^{\mathcal{N}=3} \delta[\eta_{\nu} - (\boldsymbol{\Gamma}_{\alpha} - \boldsymbol{\kappa})\cdot\mathbf{a}_{\mathrm{p}}^{\nu}],\qquad(\text{II.6})$$

$$\Psi_{\kappa}^{\gamma\,*}(\mathbf{r}_0) = \sum_{\lambda} \tilde{\Phi}_{\lambda}^{\gamma} \exp[-i(\boldsymbol{\Gamma}_{\lambda} - \boldsymbol{\kappa})\cdot\mathbf{r}_0].\qquad(\text{II.7})$$

By multiplication both sides of (II.5):

$$\begin{aligned}
1 = & (2\pi)^3 \sum_{\gamma\in\mathbb{N}}\sum_{\alpha}\Big[\frac{1}{\xi_{\gamma}}\exp(i\boldsymbol{\eta}\cdot\mathbf{r}_0)|\boldsymbol{\eta}|^2 \tilde{\Phi}_{\alpha}^{\gamma}\Psi_{\kappa}^{\gamma\,*}(\mathbf{r}_0)\prod_{\nu=1}^{\mathcal{N}=3}\delta[\eta_{\nu}-(\boldsymbol{\Gamma}_{\alpha}-\boldsymbol{\kappa})\cdot\mathbf{a}_{\mathrm{p}}^{\nu}]\Big]_{\boldsymbol{\eta}=\boldsymbol{\Gamma}_{\alpha}-\boldsymbol{\kappa}} \\
& + (2\pi)^3 \sum_{\gamma\in\mathbb{N}}\sum_{\alpha}\Big[\frac{1}{\xi_{\gamma}}\exp(i\boldsymbol{\eta}\cdot\mathbf{r}_0)\sum_{\beta}\sum_{j=1}^{\mathcal{L}}\sum_{ns=1}^{N}\tilde{\Phi}_{\alpha}^{\gamma}\Psi_{\kappa}^{\gamma\,*}(\mathbf{r}_0)C'_{ns}\exp(-i\boldsymbol{\eta}\cdot\hat{\mathbf{r}}_{ns})\prod_{\nu=1}^{\mathcal{N}=3}\delta[\eta_{\nu}-(\boldsymbol{\Gamma}_{\alpha}-\boldsymbol{\kappa})\cdot\mathbf{a}_{\mathrm{p}}^{\nu}]\Big]_{\boldsymbol{\eta}=\boldsymbol{\Gamma}_{\alpha}-\boldsymbol{\kappa}}.
\end{aligned}\qquad(\text{II.8})$$

The condition $\boldsymbol{\eta}=\boldsymbol{\Gamma}_{\alpha}-\boldsymbol{\kappa}$ arises from the fact that we have delta functions of $\boldsymbol{\eta}$ on the right hand side. From the completeness criterion we have

$$\delta^{(3)}(\mathbf{r}) = \sum_{\gamma'\in\mathbb{N}} \Psi_{\kappa}^{\gamma'}(\mathbf{r})\,\Psi_{\kappa}^{\gamma'\,*}(0).\qquad(\text{II.9})$$

This may give us an idea to state a delta-based expression instead of the unity on the left hand side of (II.8):

$$1 = (2\pi)^3 \sum_{\gamma'\in\mathbb{N}}\sum_{\alpha'}\Big[\tilde{\Phi}_{\alpha'}^{\gamma'}\Psi_{\kappa}^{\gamma'\,*}(0)\prod_{\nu=1}^{\mathcal{N}=3}\delta[\eta_{\nu}-(\boldsymbol{\Gamma}_{\alpha'}-\boldsymbol{\kappa})\cdot\mathbf{a}_{\mathrm{p}}^{\nu}]\Big].\qquad(\text{II.10})$$

Substitution of (II.10) into (II.8) and equating the weights of delta functions on both sides, results in:

$$\sum_{\boldsymbol{\alpha}'} \sum_{\gamma' \in \mathbb{N}} [\tilde{\Phi}_{\boldsymbol{\alpha}'}^{\gamma'} \Psi_{\boldsymbol{\kappa}}^{\gamma'^{*}}(0) \prod_{\nu=1}^{N=3} \delta[\eta_{\nu} - (\boldsymbol{\Gamma}_{\boldsymbol{\alpha}'} - \boldsymbol{\kappa}).\mathbf{a}_{p}^{\nu}]] =$$

$$\sum_{\boldsymbol{\alpha}} \sum_{\gamma \in \mathbb{N}} \{[\frac{1}{\xi_{\gamma}} \exp[i(\boldsymbol{\Gamma}_{\boldsymbol{\alpha}} - \boldsymbol{\kappa}).\mathbf{r}_{0}] \tilde{\Phi}_{\boldsymbol{\alpha}}^{\gamma} \Psi_{\boldsymbol{\kappa}}^{\gamma^{*}}(\mathbf{r}_{0})[|\boldsymbol{\Gamma}_{\boldsymbol{\alpha}} - \boldsymbol{\kappa}|^{2} + \sum_{\boldsymbol{\beta}}^{\mathcal{L}} \sum_{j=1}^{N} \sum_{ns=1}^{N} C'_{ns} \exp[-i(\boldsymbol{\Gamma}_{\boldsymbol{\alpha}} - \boldsymbol{\kappa}).\hat{\mathbf{r}}_{ns}]] \prod_{\nu=1}^{N=3} \delta[\eta_{\nu} - (\boldsymbol{\Gamma}_{\boldsymbol{\alpha}} - \boldsymbol{\kappa}).\mathbf{a}_{p}^{\nu}]\},$$

(II.11)

and

$$\sum_{\gamma' \in \mathbb{N}} \tilde{\Phi}_{\boldsymbol{\alpha}}^{\gamma'} \Psi_{\boldsymbol{\kappa}}^{\gamma'^{*}}(0) = \sum_{\gamma \in \mathbb{N}} \{\frac{1}{\xi_{\gamma}} \tilde{\Phi}_{\boldsymbol{\alpha}}^{\gamma} \Psi_{\boldsymbol{\kappa}}^{\gamma^{*}}(\mathbf{r}_{0}) \exp[i(\boldsymbol{\Gamma}_{\boldsymbol{\alpha}} - \boldsymbol{\kappa}).\mathbf{r}_{0}][|\boldsymbol{\Gamma}_{\boldsymbol{\alpha}} - \boldsymbol{\kappa}|^{2} + \sum_{\boldsymbol{\beta}}^{\mathcal{L}} \sum_{j=1}^{N} \sum_{ns=1}^{N} C'_{ns} \exp[-i(\boldsymbol{\Gamma}_{\boldsymbol{\alpha}} - \boldsymbol{\kappa}).\hat{\mathbf{r}}_{ns}]]\}. \quad (II.12)$$

where $m, n, p, m_1, n_1, p_1 \in [-\mathcal{T}, \mathcal{T}]$.

Employing (II.12) would be possible when all eigenvectors and eigenvalues are determined with an iterative method (or any efficient approach) from (32). Notice that the summation sweeps over all eigenstates and eigenvalues for any choice of $m$, $n$ or $p$. In fact for any variation of $m$ or $n$ or $p$ we obtain a linear equation as $f(c_e^1, c_e^2, ..., c_e^N, \boldsymbol{\kappa}) = 0$. Obviously, various combinations of $(m, n, p)$ give rise to $(2\mathcal{T}+1)^3$ linear equations, which may be inefficient for estimating all $c_e^{ns}$ if $N > (2\mathcal{T}+1)^3$, with a given $\boldsymbol{\kappa}$. However, at the time of calculating eigenvectors (without any error assumption), numerous values of wavevectors enter as the input (through sweeping paths in $\boldsymbol{\kappa}$-space). Based on the desired accuracy for plotting curves, the number of inputs $N_{\boldsymbol{\kappa}}$ could change. So a system of equations is obtained, which relates the wavevectors and their relevant eigenvalues through

$$\begin{pmatrix} f(c_e^1, c_e^2, ..., c_e^N, \boldsymbol{\kappa}_1) \\ f(c_e^1, c_e^2, ..., c_e^N, \boldsymbol{\kappa}_2) \\ \vdots \\ f(c_e^1, c_e^2, ..., c_e^N, \boldsymbol{\kappa}_N) \end{pmatrix} = \begin{pmatrix} 0 \\ 0 \\ \vdots \\ 0 \end{pmatrix}, N_{\boldsymbol{\kappa}} \geq N. \quad (II.13)$$

Resultant errors $(c_e^1, c_e^2, ..., c_e^N)$ of solving the above linear system produces new sample weights, which can be entered in an iterative procedure for determination of eigenvectors and eigenvalues. In practice, an accurate eigenvalue extraction leads to small errors even at initial steps; hence, small aberrations would vanish gradually.

**Table Captions**

Table 1. Comparison between energies of high symmetry points with the non-local pseudopotential model (including spin-orbital interaction) and sampling method.

Table. 2. Comparison between energy values of high symmetry points attained from ab initio and sampling methods. AF=0.38, $\mathcal{C}=10$ and *ns*=540.

**Figure Captions**

Fig. 1.  Spatial location of samples (*ns*=125) and discrete potential distribution (delta weights) are shown for a Germanium closed shell with lattice constant of 5.66Å along principal coordinates. AF=0.1, N =4 and *ncn*=24 (potential is in the range of *meV*). Maxima of X, Y and Z equals to $N/(N+1).AF.a$ or approximately 0.453Å, while size of intervals is $AF.a/(N+1)$ or 0.1132 Å.

Fig. 2.  (a) Spatial profile for potential of Ge closed shell containing three z contours. AF=0.1, N=10, *ns*=1331, *ncn*=24 (potential is in the range of meV). (b) Changes of S-domain, resolution and interval with respect to AF reduction. For an AF greater than 0.25, S-domain overshoots primitive cell boundaries.

Fig. 3.  Eigenvectors relevant to the first five energy bands (dots) for $\mathcal{C}=2$, AF=0.1, *ns*=1331, *ncn*=24. A symmetric distribution would be obtained if we shift the chart horizontally by −62. Coefficients less than 20 (and over than 100) are close to zero.

Fig. 4.  Band structure of Ge with revealed symmetry points. F=0.2 (optimized adjustment factor), $\mathcal{C}=3$, *ns*=2662, and 7 neighbors other than unit cell's atoms are considered to be included in the multipole potential.

Fig. 5.  (a) Ab-initio model for local pseudopotential. Solid line: $V_{ps}(r)$ which is utilized in OF-DFT method, with $t_{\rho KS}=0.1$, $d_0=0.02942$, $d_1=0.3119$, $d_2=0.06861$, $r_c=1.861$ and $F_1=F_2=6$; horizontal axis (distance) is normalized to Bohr radius and potential is in the range of meV. Dot-dash line: model used for sampling here. Inset: valence charge density obtained from low-q (LQ) and high-q (HQ) methods (see ref [40]). (b) distribution of delta weights with partially non-uniform sampling.

Fig. 6.  Si energy bands with revealed symmetry points. Sampling is performed on AILPS model. $\mathcal{C}=3$, *ns*=6776 and S-domain is a sphere with the radius equal to $9r_c$. 7 neighbors other than unit cell's atoms were regarded.

Fig. 7.   (a) Planar Graphene lattice with depicted primitive cell (S-domain), primitive vectors and affecting first and second order neighbors (green color for $A_1$ and blue color for $A_2$). (b) Graphene reciprocal lattice. First Brillouin zone (hexagon) is distinguished. Reciprocal basis vectors and symmetry points are shown.

Fig. 8.   Excess phase produced by $(\Gamma_\beta - \Gamma_\alpha).\mathbf{R}_p^{ns}$ for a radial sampling issue; *ns* is tried to be relatively similar for both F. It is seen that the maximum phase for AF=0.5 is larger than the same for AF=0.12. Furthermore, for AF=0.12 the mean value of entire phases is smaller. Horizontal number depends on the truncation number and the total number of samples in each cell.

Fig. 9.   (a) Graphene $\pi$ bands are shown for a slight variation of applied potential (SF=0.8, 1 and 1.25), $\mathcal{T}=10$, *ns*=540, *ncn*=9 and AF=0.38. Values of both conduction and valence bands at high symmetry points are specified. (b) Potential profile versus *r* in polar system for 0.8, 1 and 1.25 of $V_{app}$, angle=0 (black curves), and angle=$\pi/3$ (blue curve). Re-falling of black curves for *r* >0.8Å is due to the vicinity of first neighbor on $(a/\sqrt{3},0)$ (potential is in the range of 10meV). (c) Real part of eigenvectors relevant to $\pi$ bands (left: conduction, right: valence) and $\kappa = 0$. Imaginary part appears in a similar distribution of collapsing.

Fig. 10.   (a) Conduction and valence bands are depicted for two values of AF=0.33 and 0.38, while the SF, truncation (T=10 for black curves) and sample intervals are kept constant. We distinguish a slight variation in AF (and consequently in excess phase and sample numbers) affects eigenvalues enormously. Also apparently increasing truncation is not responsive over $\mathcal{T}=10$. (b) Potential profile for $V_{app}$ and various angles $(a/\sqrt{3},0)$ (potential is in the range of 10meV). (c) Variation in real part of secondary eigenvector due to a slight change (0.05) in AF (also see Fig. 9c). (d) Excess phase for AF=0.33 and 0.38.

Table 1

| S-Point | N-L Pseudopotential (eV) | Sampling method (eV) | $\Delta E$(eV) |
|---|---|---|---|
| $\Gamma_{6v}$ | -12.66 | -12.14 | -0.26 |
| $\Gamma_{7v}$ | -0.29 | -0.11 | -0.18 |
| $\Gamma_{8v}$ | 0.00 | 0.00 | 0.00 |
| $\Gamma_{7c}$ | 0.90 | 0.74 | 0.16 |
| $\Gamma_{6c}$ | 3.01 | 1.92 | 1.09 |
| $X_{5v}$ | -8.56 | -8.32 | -0.24 |
| $X_{5v}$ | -3.29 | -3.55 | 0.26 |
| $X_{5c}$ | 1.26 | 0.91 | 0.35 |
| $L_{6v}$ | -10.39 | -10.31 | -0.08 |
| $L_{6v}$ | -7.61 | -7.24 | -0.34 |
| $L_{6v}$ | -1.63 | -1.68 | 0.05 |
| $L_{4,5v}$ | -1.43 | -1.63 | 0.20 |
| $L_{6c}$ | 0.76 | 0.64 | 0.12 |
| $L_{6c}$ | 4.16 | 3.07 | 1.09 |

Table 2

| S-Point | Ab-initio | Sampling | $\Delta E$(eV) |
|---------|-----------|----------|----------------|
| $\Gamma_v$ | -7.6 | -6.36 | -1.24 |
| $\Gamma_c$ | 11.3 | 11.28 | 0.02 |
| $M_c$ | -2.3 | -2.56 | 0.26 |
| $M_v$ | 1.7 | 2.98 | -1.26 |

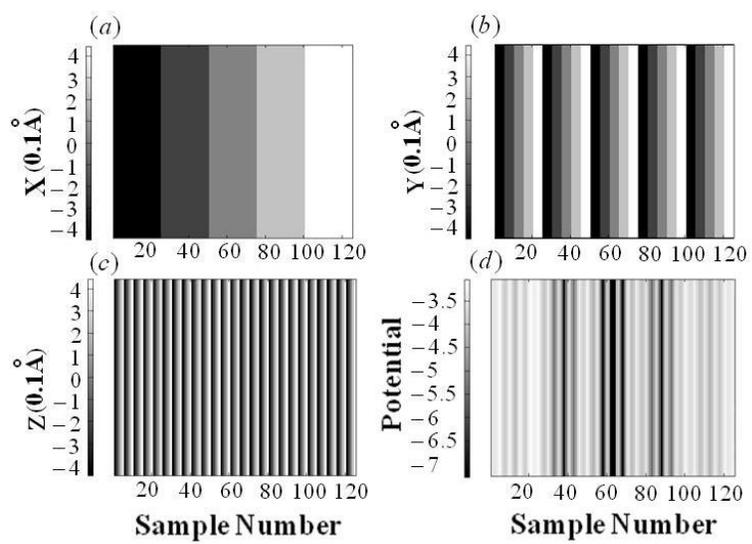

Fig. 1.

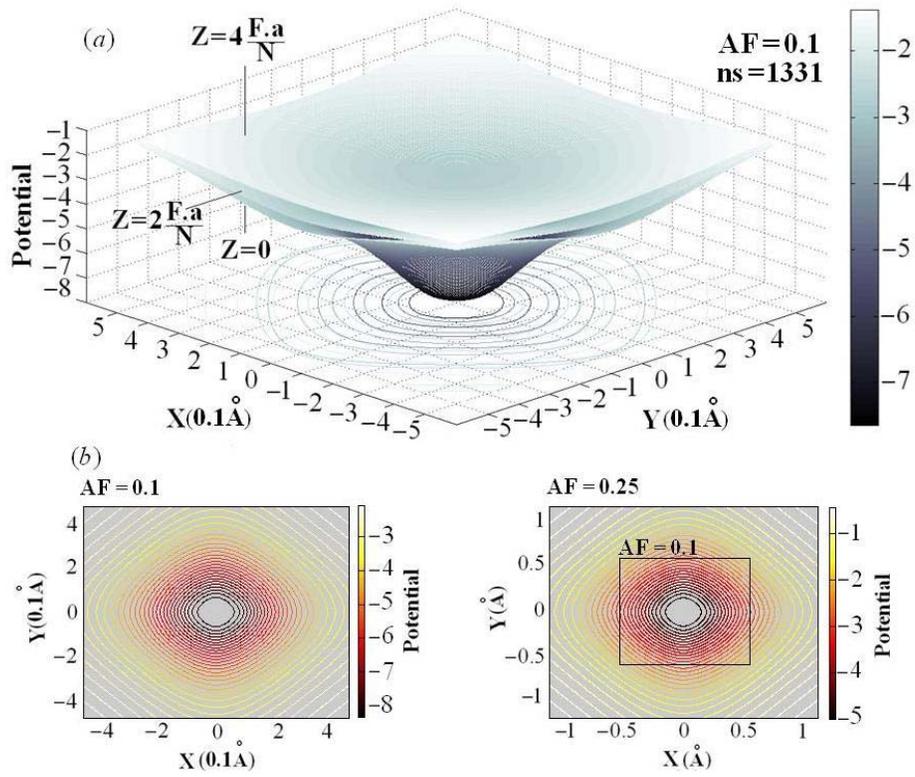

Fig. 2.

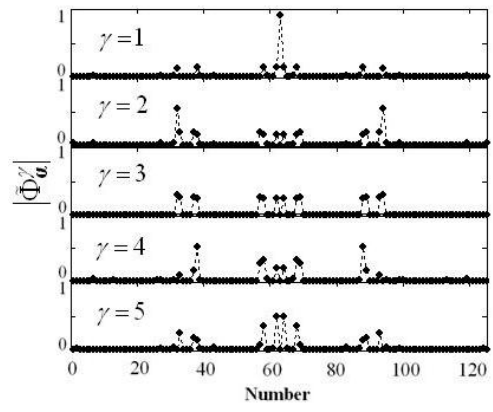

Fig. 3.

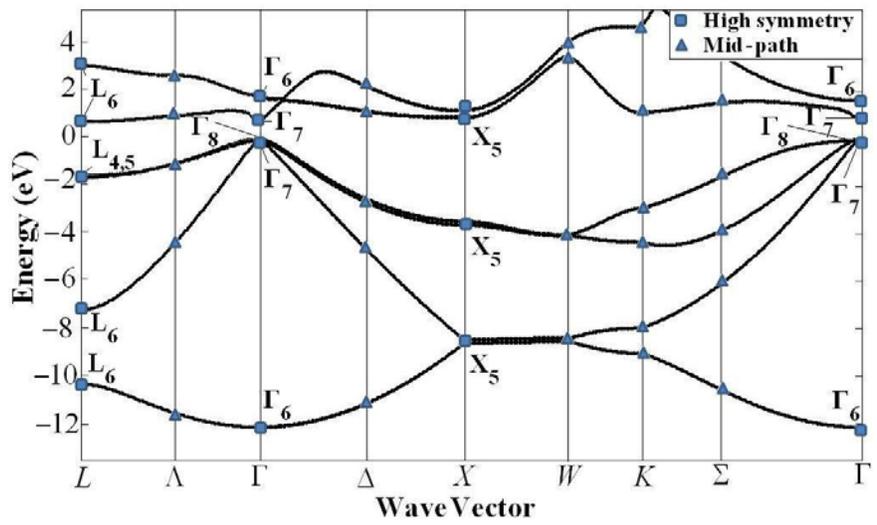

Fig. 4.

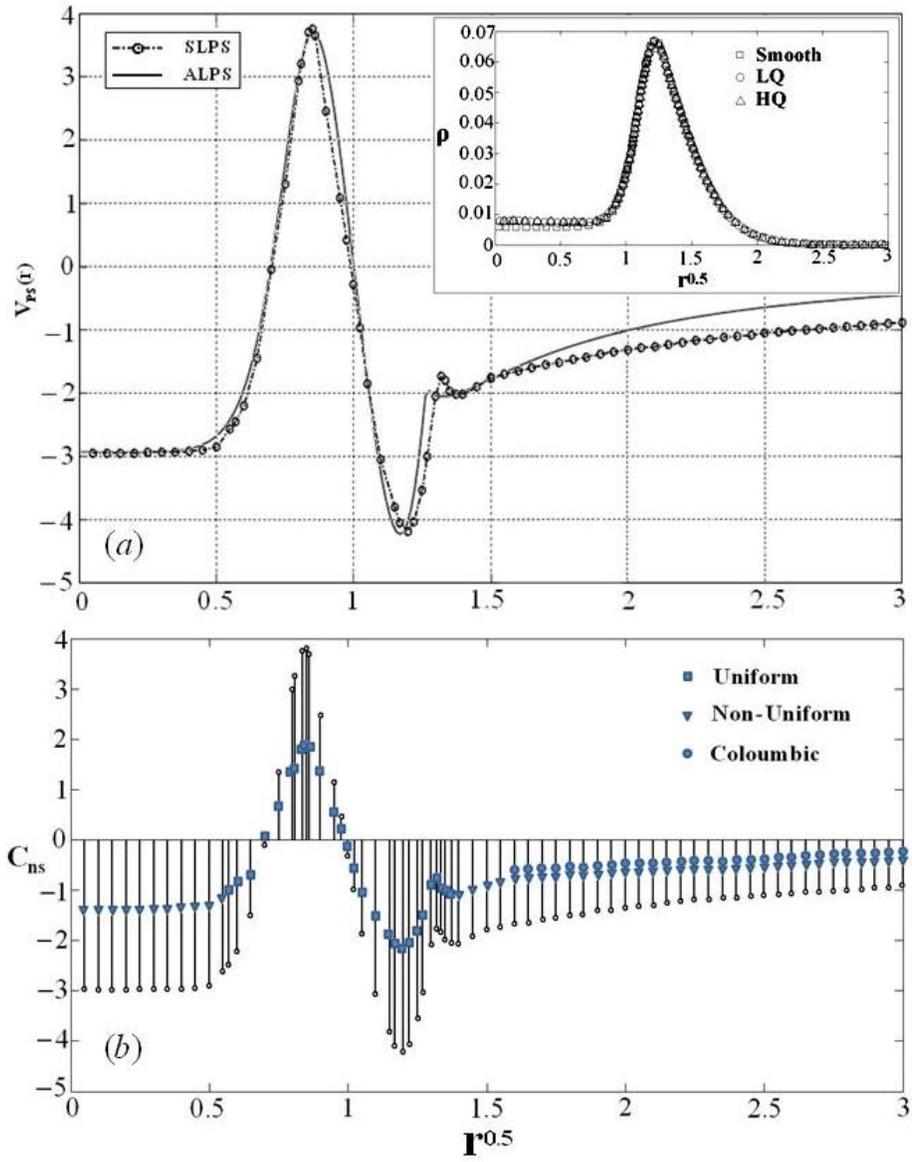

Fig. 5.

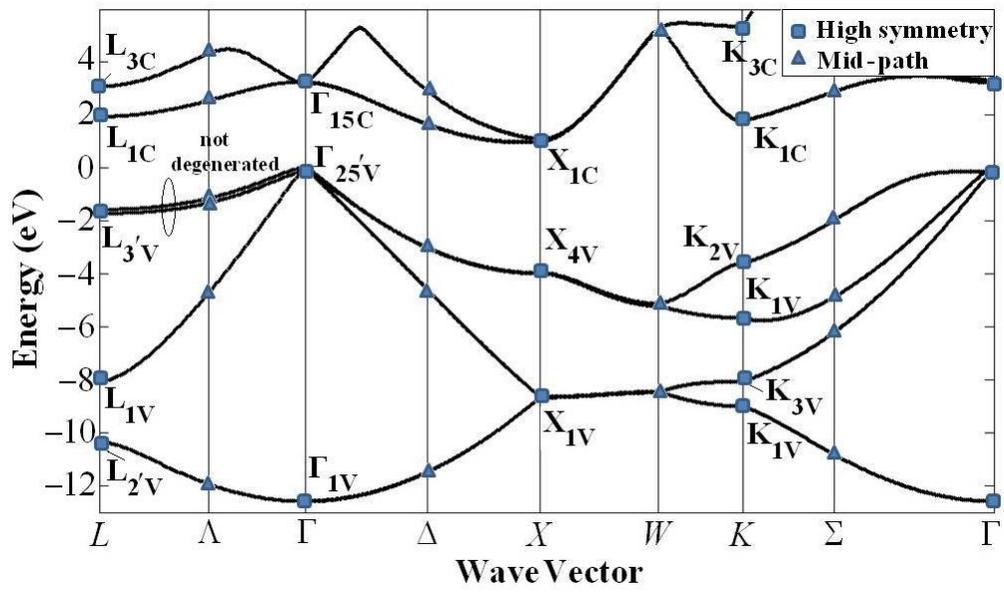

Fig. 6.

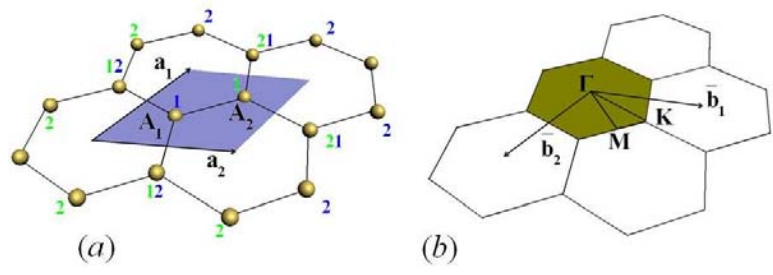

Fig. 7.

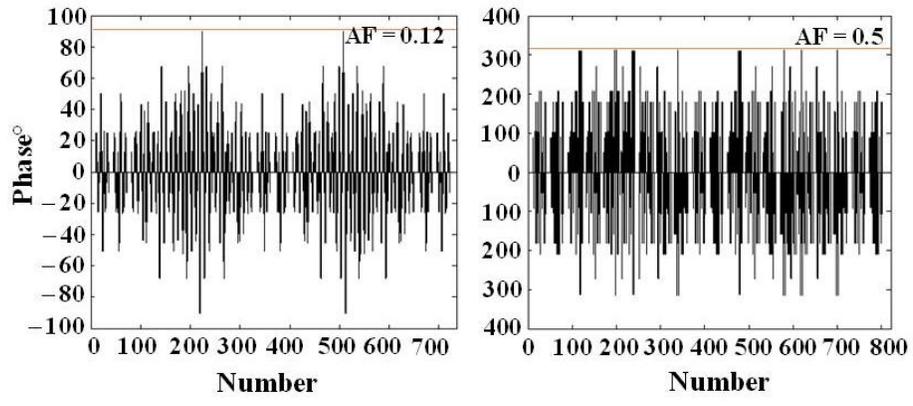

Fig. 8.

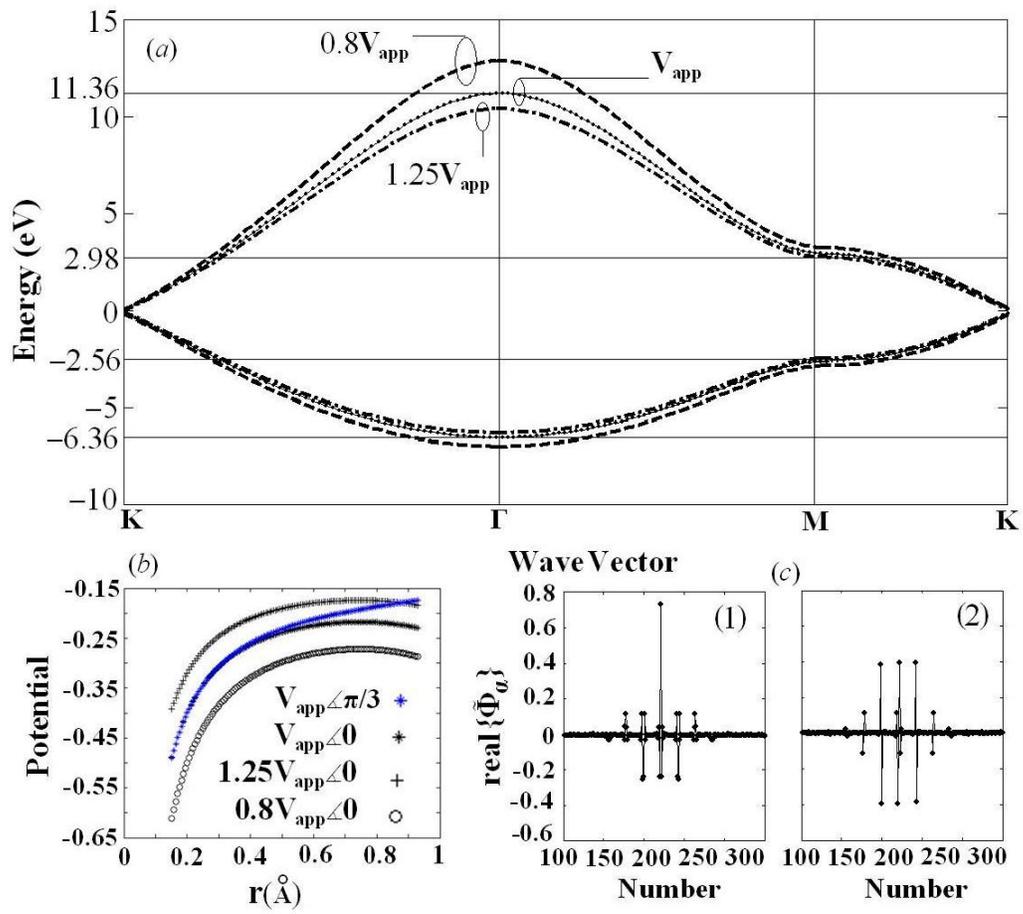

Fig. 9.

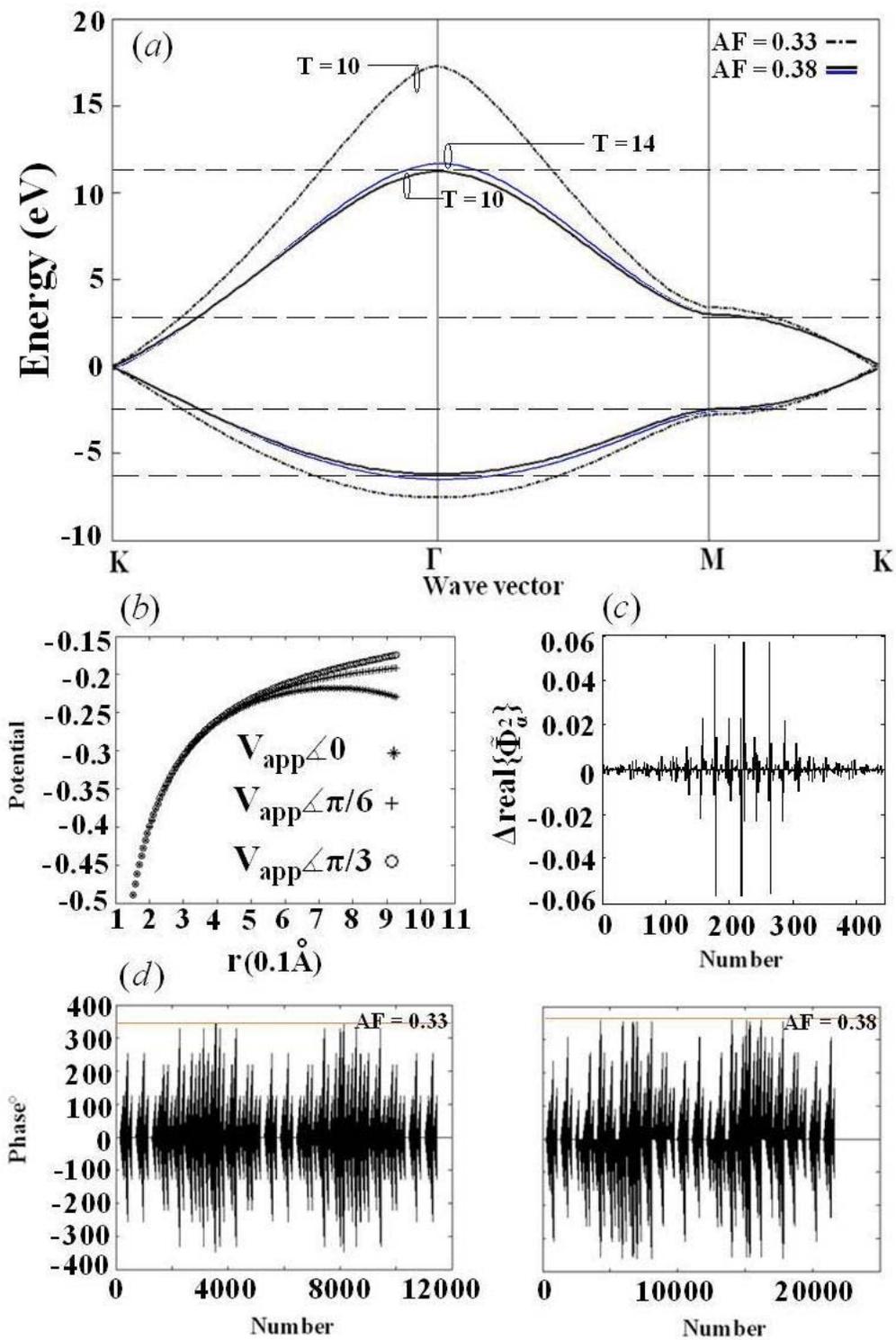

Fig. 10.